\documentclass[prl,twocolumn,showpacs,preprintnumbers,amsmath,amssymb]{revtex4}

\usepackage{graphicx}% Include figure files

\usepackage{graphicx}% Include figure files
\usepackage{xcolor}

\begin{document}

%\title{Ultrafast light-induced shear strain probed by time-resolved X-ray diffraction}
\title{Ultrafast light-induced shear strain probed by time-resolved X-ray diffraction:  the model multiferroic BiFeO$_3$ as a case study}

\author{V. Juv\'e $^{1}$, R. Gu$^{1}$, S. Gable$^{2}$, T. Maroutian$^{2}$, G. Vaudel$^{1}$, S. Matzen$^{2}$, N. Chigarev$^{3}$, S. Raetz$^{3}$, V. E.  Gusev$^{3}$, M. Viret$^{4}$, A. Jarnac$^{5}$, C. Laulh\'e $^{5}$, A. Maznev$^{6}$, B. Dkhil$^{7}$, P. Ruello$^{1}$\footnote{ Electronic address: pascal.ruello@univ-lemans.fr}}

\affiliation{
$^{1}$Institut des Mol\'ecules et Mat\'eriaux du Mans, UMR 6283 CNRS, Le Mans Universit\'e, 72085 Le Mans,  France\\
$^{2}$ Centre de NanoSciences et de Nanotechnologies, UMR 9001 CNRS Universit\'e Paris Saclay, Palaiseau, France.\\
$^{3}$Laboratoire d'Acoustique de Le Mans Universit\'e, UMR CNRS 6613, Le Mans Universit\'e, 72085 Le Mans, France\\
$^{4}$ SPEC UMR CEA/CNRS, Universit\'e Paris-Saclay, L'Orme les Merisiers, 91191 Cedex, Gif-sur-Yvette, France\\
$^{5}$ Université Paris-Saclay, Synchrotron Soleil, 91190, Saint-Aubin, France.\\
$^{6}$ Department of Chemistry, Massachusetts Institute of Technology, Cambridge, MA 02139, USA \\
$^{7}$Laboratoire Structures, Propri\'et\'es et  Mod\'elisation des Solides, CentraleSup\'elec, UMR CNRS 8580, Universit\'e Paris-Saclay, 91190 Gif-sur-Yvette, France}

\begin{abstract}
Enabling the light-control of complex systems on ultra-short timescales gives rise to rich physics with promising applications. While crucial, the quantitative determination of both the longitudinal and  shear photo-induced strains still remains challenging. Here, by scrutinizing asymmetric Bragg peaks pairs $(\pm h01)$ using picosecond time-resolved X-ray diffraction experiments in BiFeO$_3$, we simultaneously determine the longitudinal and shear strains. The relative amplitude of those strains can be explained only if both thermal and non-thermal processes contribute to the acoustic phonon photogeneration process. Importantly, we also reveal a difference of the dynamical response of the longitudinal strain with respect to the shear one due to an interplay of quasi-longitudinal and quasi-transverse acoustic modes, well reproduced by our model. 

\end{abstract}

\maketitle

In principal coordinate system, the longitudinal strain in matter is represented by a compressional field with diagonal tensor. On the contrary, the shear strain tensor has only off-diagonal terms associated with a zero volume change. Specifically, it is the $curl$ part (rotational motion) of the atomic displacement which propagates at the shear velocity in matter~\cite{meca}, i.e. shear acoustic phonons carry angular momentum. In the context of ultrafast science in condensed matter, this specific symmetry of atomic displacements has several impacts: for examples the light-induced picosecond shear pulse can be used to probe ultrafast friction in soft matter~\cite{pezeril2009,klieber} or induce ultrafast rotation of light-polarization  ~\cite{matsuda2004,pezeril2007,matsuda2008,mounier}. It has been also shown that during an ultrafast light-induced demagnetization process the shear acoustic phonons can exchange angular momentum with spins through the Einstein-de Haas effect~\cite{einstein, johnson} or sometimes called the Richardson effect~\cite{richardson,wern}. Beyond these two examples, generating shear motion with light has received a great deal of attention in general and demonstration of this phenomenon has been reported in different materials like multiferroic oxides such as BiFeO$_3$~\cite{ruello2012,lejman2014,lejman2019}, piezoelectric semiconductors GaN or GaAs~\cite{wen,matsuda2004}, metals~\cite{pezeril2007,matsuda2004,ruelloLU,gusev1993} or spin-crossover compounds~\cite{parpiiev} for citing a few. Despite this active and continuous effort, the underlying physics of the light-induced shear strain generation remain unclear since the quantitative measurement of the shear strain amplitude is lacking. X-ray or electron diffraction methods appear as the natural experimental tools to quantify the light-induced strain. However time-resolved X-ray~\cite{lindenberg,reis,trigo,larsson,korff2007,schick,johnson,lee,wenPb2013,daranciang2013,matzen,venka} or electron~\cite{arbouet,feist,naka} diffraction experiments have mostly been applied to extract the longitudinal strain. A few attempts of a quantitative evaluation of the light-induced shear strain amplitude have been recently reported in crystalline organic thin film by time-resolved X-ray diffraction with 100~ps of resolution~\cite{cama} or in very thin layers, either by analysing the time-dependence of intensity of electron diffraction peak in layers of VTe$_2$~\cite{naka} or by analysing the Bragg peak shifts in graphite~\cite{feist}.

In this Letter we apply picosecond time-resolved X-ray diffraction and measure the transient evolution of the asymmetric Bragg reflection pairs $(\pm h01)$ to quantitatively determine the photoinduced shear and longitudinal strains in a BiFeO$_3$ (BFO) single crystal unit cell. %The measurement of the transient evolution of the asymmetric Bragg reflection pairs $(\pm h01)$ quantitatively provides the shear and longitudinal strains in the unit cell. 
We evidence different temporal behavior between the shear and longitudinal strains indicating that the BFO's unit cell starts to expand during the first tens of picoseconds after the arrival of the light pulse and only then undergoes a shear deformation. This peculiar dynamic is well reproduced by our modeling based on strains wave propagation theory. Moreover, quantitative information on the strain amplitudes sheds new light on the photoinduced generation process and indicates that both thermal and non-thermal processes are at play. Beyond the BFO case, our work shows that this approach is very versatile and can be employed to access to the in-plane atomic displacements in any kind of structure where an in-plane symmetry breaking is present or initiated by an external stimulus including light pulses.
%\columnwidth
\begin{figure}[!t]
\centerline{\includegraphics[width=\columnwidth]{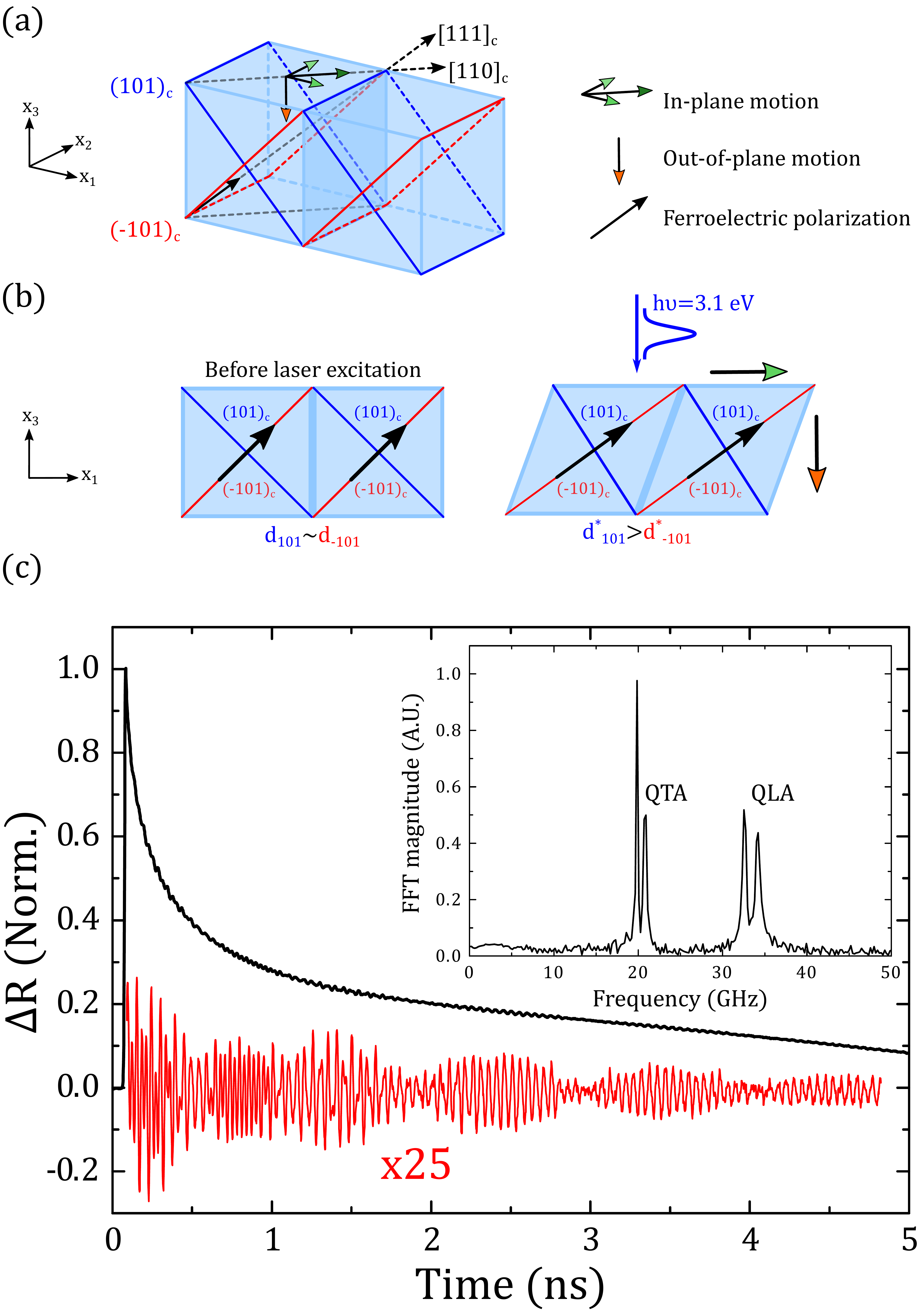}}
\caption{\label{fig1}
 (a) 3D view of the BFO pseudo-cubic unit cell with displacement vector associated to the longitudinal and shear photoinduced strain (orange and green arrows). The black arrow is the BiFeO$_3$ ferroelectric polarization pointing along the $[111]_{c}$ direction. (b) Side view of the BiFeO$_3$ pseudo-cubic unit cell before and after the laser excitation showing the in-plane and out-of-plane motion. (c) Transient optical reflectivity signal of BFO single crystal measured with a 400~nm pump and a 800~nm probe beams. The red line shows the acoustic phonon signal once the base line has been subtracted. The inset shows the fast Fourier transform with the quasi-longitudinal and quasi-transverse acoustic phonon modes.}
\end{figure}

We first describe theoretically how the unit cell is distorted in presence of a longitudinal and shear strain. Knowing the rhombohedral distortion in bulk BFO is weak (equivalent rhombohedral angle $\sim89.5^{\circ}$, and pseudocubic parameter $a_0\sim$3.96 $\mathrm{\AA}$), we consider the pseudo-cubic representation of the BFO lattice as depicted in Fig~\ref{fig1}(a). At equilibrium (before laser excitation), the interplanar distances of the $(101)_{c}$ and $(-101)_{c}$ planes are nearly identical ($d_{101}\approx d_{-101}$). As the laser pulse impinges on the $(001)_{c}$ surface of BFO and due to the existence of in-plane symmetry breaking caused by the ferroelectric polar order, the laser-matter interaction leads to the generation of shear motion  ~\cite{ruello2012,lejman2014,lejman2016,lejman2019} in addition to the longitudinal strain. Importantly, the shear strain is expected to lead to an asymmetric change of the interplanar distances $d^\ast_{101}$ and $d^\ast_{-101}$ with $d^\ast_{101}>d^\ast_{-101}$ as displayed in Fig.~\ref{fig1}(b). This principle can be applied to all ($\pm h01$) planes. Note that the atomic displacements associated to the longitudinal and shear strain are symbolized with orange and green arrows respectively in Figs.~\ref{fig1}(a,b). The relation between the interplanar distance $d^\ast_{\pm h01}$ and the longitudinal $\eta_L$ and shear $\eta_S$ strains is established in the Supplementary Note 2 as: 
\begin{eqnarray}
\label{dbragg1}
d^\ast_{\pm h01}&=&d_{\pm h01}\bigg (1+ \frac{\eta_L \pm |h|\frac{\eta_{S}}{\sqrt{2}}}{1+h^2}\bigg).
\end{eqnarray}

Measuring both $d^*_{h01}$ and $d^*_{-h01}$ then allows disentangling the photoinduced longitudinal and shear strains. They cast as:
\begin{equation}
\label{strainLS}
\begin{aligned}
\eta_L&=&\frac{1+h^2}{2} \times \bigg[{\frac{d^\ast_{h01}-d_{h01}}{d_{h01}}+\frac{d^\ast_{-h01}-d_{-h01}}{d_{-h01}}}\bigg] \\
\eta_{S}&=&\frac{1+h^2}{\sqrt{2}\mid h\mid} \times \bigg[{\frac{d^\ast_{h01}-d_{h01}}{d_{h01}}-\frac{d^\ast_{-h01}-d_{-h01}}{d_{-h01}}}\bigg] .
  \end{aligned}
\end{equation}

The experiments make use of a pump-probe scheme where a femtosecond pulse with a photon energy of 3.1~eV higher than the band gap (E$_g\sim 2.6$~eV) generates strain pulses in a single BFO crystal oriented along the $[001]_{c}$ direction~\cite{BFOmichel}. The propagation of the strain pulses is then followed in the crystal's depth either by another femtosecond pulse with a photon energy of 1.5~eV or by a synchronized hard X-ray picosecond pulse. In the all-optical pump-probe experiment, detection with a photon energy below the band gap allows detecting the emitted coherent acoustic phonons over a long time of propagation by means of Brillouin scattering. In Fig.~\ref{fig1}(c), the change of the probe's reflectively is plotted as a function of the pump-probe delay. Once the slow relaxation signal is subtracted, the coherent acoustic phonons' signal can be isolated (red curve in Fig.~\ref{fig1}(c)) and its spectrum extracted with a fast Fourier transform (inset of Fig.~\ref{fig1}(c)). In the case of BFO, the existence of the symmetry plane $(110)_{c}$ being perpendicular to the irradiated surface, restricts the light-induced atomic motions within this $(110)_{c}$ plane and consequently suppresses the pure shear motion. %perpendicular to this plane by symmetry principle and 
It turns out that only quasi-longitudinal (QLA) and quasi-transverse (QTA) modes are excited in agreement with previous studies~\cite{ruello2012,lejman2014}. In the detection process, the QLA and QTA modes are each split into two, as shown in Fig. \ref{fig1}(c),  due to BFO's optical birefringence as already discussed in a previous report~\cite{lejman2016}. The corresponding sound velocities are $V_{QLA}=4970 \pm 30~m.s^{-1}$ and $V_{QTA}=3020 \pm 30~m.s^{-1}$ as deduced from the Brillouin frequency equation $f_B=2n_{probe}V/\lambda_{probe}$ where $n_{probe}$ is the BFO ordinary or extraordinary refractive index at the probing wavelength $\lambda_{probe}$~\cite{choi}. %The calculated sound velocities are in agreement with previous measurements on different BFO$_3$ samples~\cite{lejman2014,ruello2012}. 

%In order to have more insight into the photoinduced strain generation at the atomic scale, we conducted picosecond time-resolved X-ray diffraction measurements on the BFO single crystal. 
\begin{figure}[!t]
\centerline{\includegraphics[width=\columnwidth]{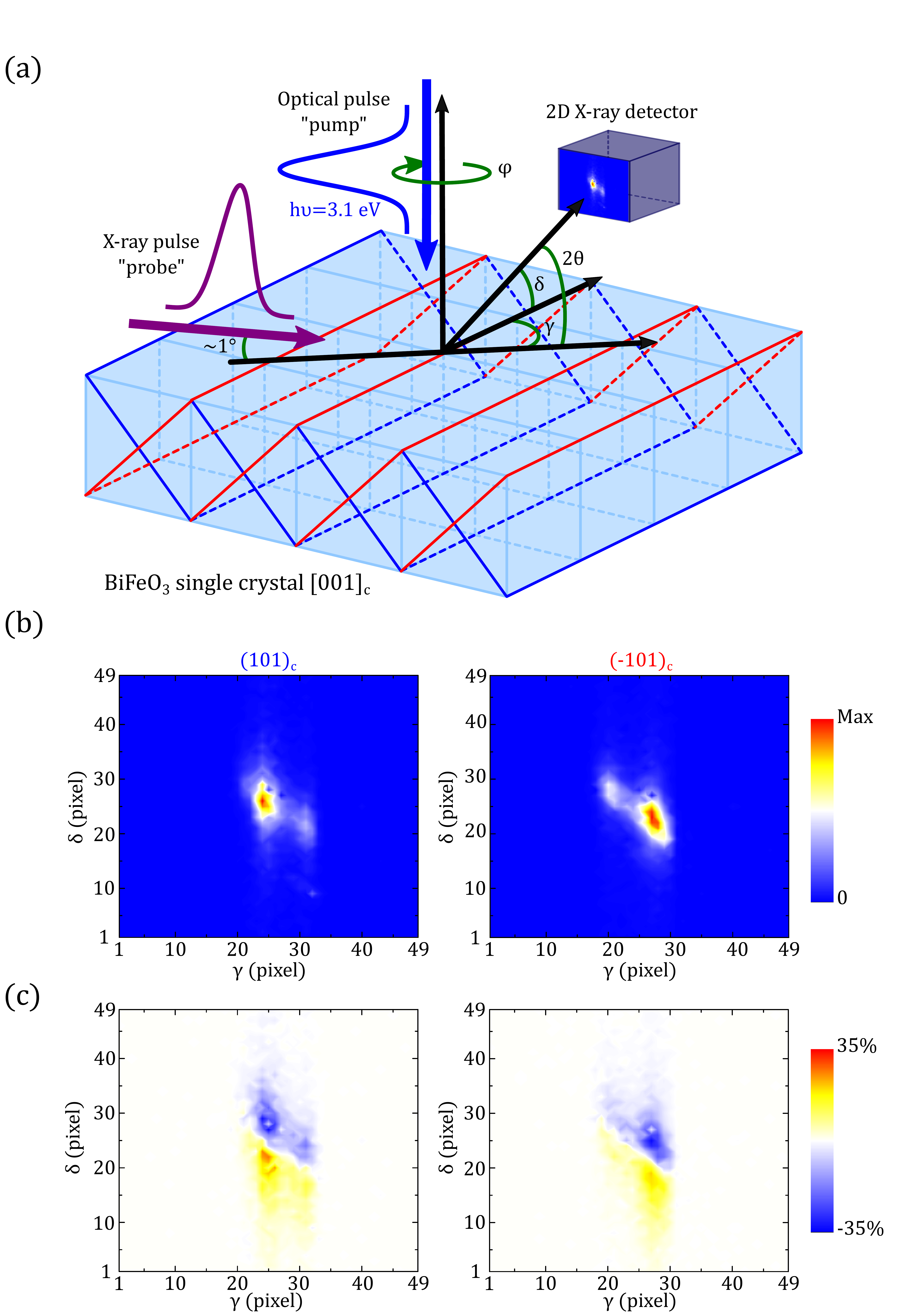}}
\caption{\label{fig2} 
(a) Scheme of the time-resolved X-ray diffraction in the grazing incidence geometry. (b) Reciprocal space imaging of the $(101)_{c}$ and $(-101)_{c}$ Bragg diffraction peaks at equilibrium. The positions ($\gamma,\delta$) of the $(101)_{c}$ and $(-101)_{c}$ Bragg diffraction peaks are $(26.1874^\circ,26.0802^\circ)$ and (26.0367$^\circ$, 25.6879$^\circ$) respectively with a camera resolution of $8.64\times 10^{-3}~^\circ/$pixel. (c) Differential reciprocal space imaging of the $(101)_{c}$ and $(-101)_{c}$ Bragg diffraction peaks at a time delay of 200~ps and at a fluence of 3~mJ.cm$^{-2}$.}
\end{figure}
Regarding time-resolved X-ray diffraction experiments, the measurements were performed at the SOLEIL Synchrotron on the CRISTAL beamline in the low-alpha mode, with hard X-ray pulses of 7.155 keV energy and 12 ps duration \cite{cristal}. The X-ray grazing incidence geometry ($1^\circ$, see Fig.~\ref{fig2}(a)) was used to match the optically excited and probed volumes. The effective penetration depths of the 3.1 eV pump beam and of the X-ray beam can be estimated to $\xi_p \sim$40~nm~\cite{choi} and $\xi_X \sim$70~nm~\cite{xab}, respectively. In this diffraction geometry, the Bragg angle $\theta$ is deduced from the relation $\cos(2\theta)=\cos(\delta)\cos(\gamma)$. The pump beam size on the sample was about twice as large as the X-ray probe beam size and the pump beam fluence was set to either 1.5 or 3~mJ.cm$^{-2}$. %The time-resolved X-ray diffraction experiments were conducted in a quasi-stationnary regime where the pulsed femtosecond laser continuously impinged on the sample. When the photoexcitation was stopped, we have observed a very slow relaxation process of the lattice with characteristic time of the order of the minute. This very slow dynamics is out of the scope of the paper where we concentrate on the ultrafast response (picosecond time scale) of the lattice after femtosecond light excitation. 
The $(\pm h01)_{c}$ Bragg peaks were recorded as the function of the pump-probe delay with gateable detector XPAD3.2 \cite{detector}. The sample was rotated by 180$^\circ$ to switch between $(h01)_{c}$ and $(-h01)_{c}$ Bragg diffraction peaks after a complete pump-probe delay scan. Few scans were recorded for each set of experimental parameters, namely $\pm h$ and the pump fluence. Fig.~\ref{fig2}(b) shows typical images of the $(101)_c$ and $(-101)_c$ Bragg peaks before laser excitation. 

Those are split, which indicates the presence of a secondary small ferroelectric-ferroelastic domain in the BFO single crystal. After laser excitation, the $(101)_c$ and $(-101)_c$ peaks evolve as depicted in Fig.~\ref{fig2}(c) for a pump-probe delay of 200~ps. The 2D center of mass ($\gamma (t),\delta (t)$) was calculated for each pump-probe delay, which allowed us to extract the time-dependence of the relative interplanar distance $\Delta d(t)/d$ after laser excitation (see Supplementary Note 3).

\begin{figure}%[t!]
\centerline{\includegraphics[width=8cm]{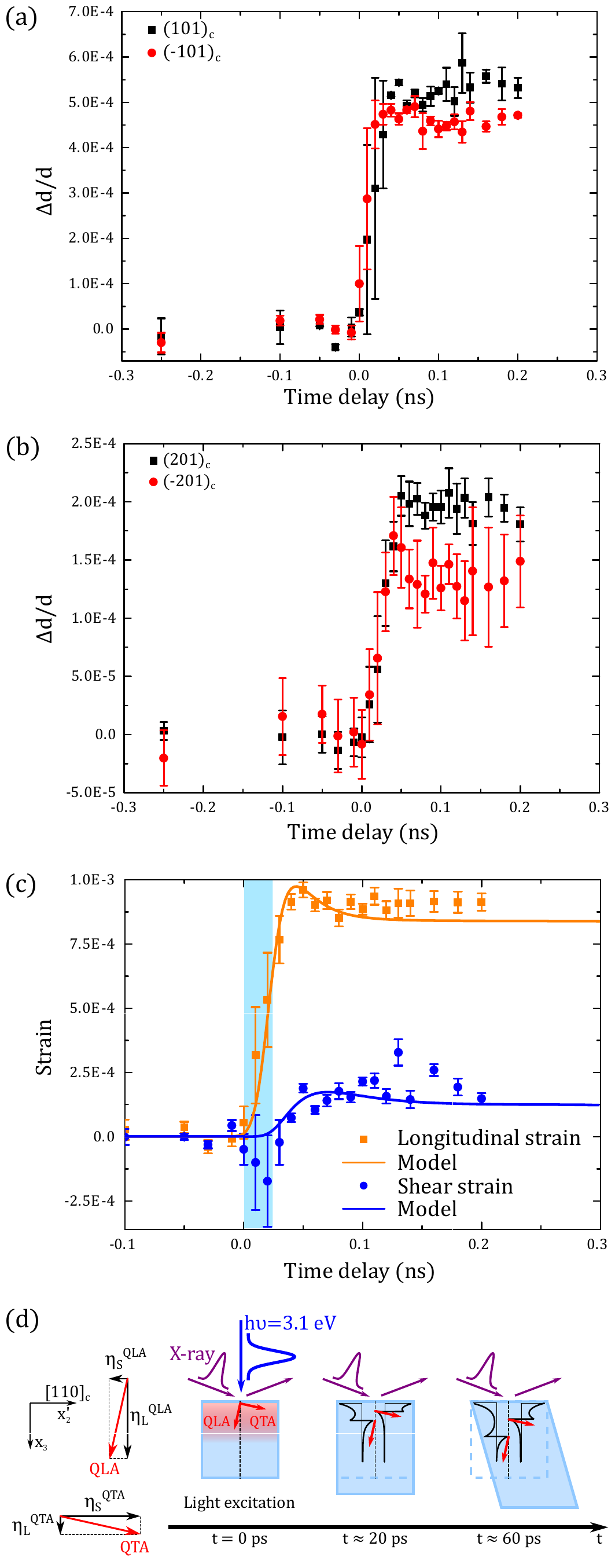}}
\caption{\label{fig3} 
(a) Time dependence of the relative variation of the interplanar distance for the $(101)_{c}$ and $(-101)_{c}$ planes. (b) Same as (a) but for the $(201)_{c}$ and $(-201)_{c}$ planes. (c) Comparison of the experimentally determined photoinduced strains (symbols) with theory (lines) as a function of the time delay. The experimental strains were calculated by using Eq.~\ref{strainLS} and the theory strains refer to  $<\eta_S>(t)$ and $<\eta_L>(t)$ from Eq.~\ref{strainLSaverage}. (d) Sketch of the time-dependent shape of the unit cell (from a cubic to a trapezoidal form) related to the spatial separation of the in-plane and out-of-plane strain components of the QLA and QTA modes.
}
\end{figure}

The results for $(101)_{c}$, $(-101)_{c}$, $(201)_{c}$ and $(-201)_{c}$ lattice planes are displayed in Fig.~\ref{fig3}(a)-(b) for the first 300~ps. As expected, one can clearly observe non-equivalent $\Delta d/d$ photoinduced dynamics for the $(h01)_{c}$ and $(-h01)_{c}$ lattice planes after time delay zero. The transient longitudinal and shear strains were derived by using Eq.~\ref{strainLS}. The results are displayed in Fig.~\ref{fig3}(c) and show a plateau-like response after roughly 100~ps for both strains with an amplitude ratio of $\eta_L/\eta_S\approx$ 6. Strikingly, one can clearly observe a time delay between the onsets of the longitudinal and shear strains (within the blue-shaded area) in Fig.~\ref{fig3}(c). The maximum of the longitudinal strain is reached at around 40~ps while that of the shear strain is only reached at around 60~ps. Such asynchronous behavior was never observed before as previous works were restricted only to the photoinduced longitudinal strain~\cite{schick,wenPb2013,daranciang2013}. Changing the laser pump fluence to  1.5~mJ.cm$^{-2}$ gives the same general temporal behavior but with half the amplitude, thus confirming the linear response of the system to the laser excitation (not shown).

The experimental data were analyzed by modeling the laser induced strain propagation in the BFO single crystal. We considered only the lattice dynamics and the pump (optical) and probe (X-ray) pulses penetration depths, neglecting the photoexcited carrier diffusion. We remind that as the trigonal axis is not aligned with the propagation axis of the acoustic modes $x_3$, the generated QLA and QTA modes have components both in-plane, along the $[110]_{c}$ direction, and out-of-plane, along the $[001]_{c}$ direction (see Fig.~\ref{fig3}(d)). Therefore, we can write the total shear and longitudinal strains as:
\begin{equation}
\label{strainLS2}
  \begin{aligned}
\eta_{S}(t,x_3)&=& \eta_S^{QTA} f(t-\frac{x_3}{V_{QTA}}) +\eta_S^{QLA} f(t-\frac{x_3}{V_{QLA}}) \\
\eta_{L}(t,x_3)&=& \eta_L^{QTA} f(t-\frac{x_3}{V_{QTA}}) +\eta_L^{QLA} f(t-\frac{x_3}{V_{QLA}}) ,
  \end{aligned}
\end{equation}
where $\eta_S^{QTA}$ ($\eta_S^{QLA}$) and $\eta_L^{QTA}$ ($\eta_L^{QLA}$) are the in-plane and out-of-plane components of the QTA (QLA) mode propagating at the velocity $V_{QTA}$ ($V_{QLA}$) and  $f(t-x_3/V_{QTA,QLA})$ is the function describing the space and time dependence of the strain associated to the QTA or the QLA modes in the framework developed by Thomsen and coworkers~\cite{tom1}. The vector components $\vec\eta_{QLA}=(-1,4.8)$, $\vec\eta_{QTA}=(4.8,1)$ and their associated velocities ($V_{QLA}=4710~m.s^{-1}$ and $V_{QTA}=2480~m.s^{-1}$) were calculated by using the rotated elastic stiffness tensor in the Christoffel framework (see Supplementary Note 4)  with the elastic constants measured and calculated from the work of Borissenko et al.~\cite{bori} and Shang et al.~\cite{shang} respectively (see Supplementary Note 4). To simulate the transient lattice distortion, we assume the kinematic approximation of X-ray diffraction to be valid. In the kinematic approximation of diffraction theory, the measured strains are average strains weighted by the transmission factor of X-rays along the diffraction path: 
\begin{equation}
\label{strainLSaverage}
<\eta_{L,S}>(t)=\frac{\int _{x_3=0}^\infty dx_3 e^{-\frac{L(x_3)}{\Lambda}}\eta_{L,S}(t,x_3)}{\int _{x_3=0}^\infty dx_3 e^{-\frac{L(x_3)}{\Lambda}}},
\end{equation}
$L(x_3)$ refers to the length travelled by the X-ray beam within the BFO sample when diffracted at the depth $x_3$. Its expression, which depends not only on $x_3$ but also on the diffraction angles, is derived in the Supplementary Note 5. The attenuation length of X-rays $\Lambda$= 4.22 $\mu m$ allows taking into account absorption of X-rays due to the photoelectric effect.

The curves in Fig.~\ref{fig3}(c) correspond to the calculated strains by using Eq.~\ref{strainLSaverage} and are in good agreement with the experimental data (squares and circles) by adjusting $\vec\eta_{QLA}$ and $\vec\eta_{QTA}$ amplitudes in order to constrain the vectors orthogonality. Our model reproduces this asynchronous effect that originates from the opposite sign of the in-plane components of the QLA and QTA modes ($\vec\eta_{QLA}(x'_2)$ and $\vec\eta_{QTA}(x'_2)$) which leads nearly to a cancellation of the total shear strain at early times ($<\approx 20~ps$). As the acoustic modes have different velocities, this effect fades away as the acoustic modes separate in space at longer times as sketched in Fig~\ref{fig3}(d).

We now discuss the physics underlying the femtosecond generation of the longitudinal and the shear strains in BFO. In absorbing and piezoelectric materials, the photoinduced strains can be initiated by thermal (thermoelasticity) or non-thermal (inverse piezoelectric effect and deformation potential) mechanisms~\cite{tom1,gusev1993,ruelloLU}. %The former is governed by heat generation due to the non-radiative recombination of the photoexcited electrons-holes pairs~\cite{tom1,gusev1993} whereas the latter is governed by macroscopic (inverse piezoelectric effect) or microscopic (interatomic deformation potential) modification of the internal electric field driven by photoexcited electrons-holes pairs~\cite{gusev1993,ruelloLU}. 
As the BFO bang gap decreases when increasing the hydrostatic pressure~\cite{lejman2014, gomez}, the longitudinal deformation potential stress is positive and should lead to an out-of-plane contraction of lattice. On the contrary, our measurements reveal an out-of-plane expansion of the lattice, ruling out a possible contribution of the deformation potential mechanism. As for the photoinduced stress driven by the inverse piezoelectric effect mechanism ($\sigma^{PE}$), we have established in a previous work the general expression of the relevant components~\cite{lejman2014} (see Supplementary Note 6): 
\begin{equation}
\label{piezo}
\begin{aligned}
\sigma_S^{PE}=\sigma_{4}^{PE}&=(d_{31}-d_{33})\cos(\theta)\sin(\theta) E  \\
\sigma_L^{PE}=\sigma_{3}^{PE}&=(d_{31}\cos^2(\theta)+d_{33}\sin^2(\theta)) E ,
\end{aligned}
\end{equation}
where $d_{31} \sim$ -30/-20 pm.V$^{-1}$~\cite{muri,sichel} and $d_{33} \sim$ 50/70 pm.V$^{-1}$~\cite{chen} are the piezoelectric coefficients, $\theta\sim 54^{\circ}$ is the angle between the surface normal and $x_3$ axis and $E$ is the internal depolarizing field. The theoretical contribution of the photoinduced thermoelastic stress $\sigma_{ij}^{TE}$ is assessed by calculating the tensorial expression $\sigma_{ij}^{TE}=C_{ijkl}\beta_{kl}\Delta T$ with the anisotropic thermal expansion ($\beta$) of BFO~\cite{polonica} and $\Delta T$ the temperature elevation. It reads (see Supplementary Note 6):
\begin{equation}
\begin{aligned}
\label{thermoelas}
{\sigma_{S}}^{TE}={\sigma_{4}}^{TE}=(C_{41}\beta_{1}+C_{42}\beta_{2}+C_{43}\beta_{3}+2C_{44}\beta_{4})\Delta T \\
{\sigma_{L}}^{TE}={\sigma_{3}}^{TE}=(C_{31}\beta_{1}+C_{32}\beta_{2}+ C_{33}\beta_{3}+2C_{34}\beta_{4})\Delta T , 
\end{aligned}
\end{equation}

From Eq.~\ref{piezo} and Eq.~\ref{thermoelas} we can estimate the ratio of the longitudinal and shear strains, while getting rid of the internal depolarizing field and the temperature elevation, which are not known exactly. We estimate that $\sigma_L^{PE}/\sigma_S^{PE}\approx 1$ and $\sigma_L^{TE}/\sigma_S^{TE}\approx 20$, while the ratio from our measurements is $\sigma_L/\sigma_S\approx {\eta_LV_{QLA}}/{\eta_SV_{QTA}} \approx$ 12. Thus, we can conclude that the strain generation mechanism is most likely taking its origin from both inverse piezoelectric and thermal effects. Thus, we can conclude that the strain generation mechanism is most likely taking its origin from both inverse piezoelectric and thermal effects. Our results evidence that the thermal effect cannot be neglected as proposed earlier~\cite{wenPb2013,lejman2014} and it would be interested to take it into account in calculations which at the moment reproduce only the contribution of the inverse piezoelectric effect~\cite{charles}. 

In conclusion, we have determined quantitatively the photoinduced longitudinal and shear strains at the picosecond time-scale in a BFO single crystal. Our theoretical modeling reproduces the experimental results and brings into prominence the interplay of the propagating quasi-longitudinal and the quasi-transverse modes on the measured strains in the unit cell. We evidenced that the BFO's unit cell initially starts to expand during the first tens of picoseconds after the arrival of the light pulse and then later, after around 20 ps, undergoes a shear deformation. Eventually, we show that the thermoelastic mechanism is most likely to play an important role in the photoinduced strains generation in BFO contrary to previous reports in the literature. Finally, our results demonstrate the strong potential of time resolved X-ray diffraction for extracting the temporal evolutions of both in-plane and out-of-plane lattice deformations, paving the way to explore any kind of materials where in-plane symmetry breaking can be modulated by an ultra-short light pulse or any other stimulus.

\textbf{Acknowledgements} : 
We acknowledge SOLEIL for provision of synchrotron radiation at CRISTAL beamline (proposal number 20181798). The authors thank the French National Research Agency (ANR) for support with the project UP-DOWN (N$^\circ$ ANR-18-CE09-0026-04). V.J., R.G., G.V. and P.R. acknowledge the International Laboratory IM-LED. The authors thank Charles Paillard, Laurent Bellaiche, Ryan Andrew Duncan, Hyun Doug Shinv and Matias Bargheer for fruitful discussions.

%\textbf{al Information}
%Supplementary Information is available in the online version of the paper. Reprints and permissions information is available online at www.nature.com/reprints.\\

%\textbf{Author contributions}. 

\newpage

\end{document}

%\title{Probing the electron-hole-acoustic phonon coupling at the picosecond time scale during the amorphous to crystalline phase transition in GeTe}  
%\title{Non-local photoinduced picosecond coherent acoustic phonon strain pulse during  the amorphous to crystalline phase transition in GeTe} 
%\title{Diffusive photoexcited carriers in the phase change material GeTe revealed by picosecond acoustics}  
%\title{Ultrafast photoexcited carriers lattice coupling in the phase change material GeTe revealed by picosecond acoustics}  
%\title{PRELEMINARY REPORT : \\
%Ultrafast light-induced shear strain probed by time-resolved X-ray diffraction }  
\title{Supplementary Material : Ultrafast light-induced shear strain probed by time-resolved X-ray diffraction: the model multiferroic BiFeO$_3$ as a case study}
%\title{Hot photoexcited carriers diffusion in the phase change material GeTe revealed by picosecond acoustics} 
%\title{Supersonic photoinduced strain in phase change material GeTe}  
%\title{Amorphous to crystalline phase transition in phase change material GeTe revealed by picosecond acoustics }  

\author{V. Juv\'e $^{1}$, R. Gu$^{1}$, S. Gable$^{2}$, T. Maroutian$^{2}$, G. Vaudel$^{1}$, S. Matzen$^{2}$, N. Chigarev$^{3}$, S. Raetz$^{3}$, V. E.  Gusev$^{3}$, M. Viret$^{4}$, A. Jarnac$^{5}$, C. Laulh\'e $^{5}$, A. Maznev$^{6}$, B. Dkhil$^{7}$, P. Ruello$^{1}$\footnote{ Electronic address: pascal.ruello@univ-lemans.fr}}

\affiliation{
$^{1}$Institut des Mol\'ecules et Mat\'eriaux du Mans, UMR 6283 CNRS, Le Mans Universit\'e, 72085 Le Mans,  France\\
$^{2}$ Centre de NanoSciences et de Nanotechnologies, UMR 9001 CNRS Universit\'e Paris Saclay, Palaiseau, France.\\
$^{3}$Laboratoire d'Acoustique de Le Mans Universit\'e, UMR CNRS 6613, Le Mans Universit\'e, 72085 Le Mans, France\\
$^{4}$ SPEC UMR CEA/CNRS, Universit\'e Paris-Saclay, L'Orme les Merisiers, 91191 Cedex, Gif-sur-Yvette, France\\
$^{5}$ Université Paris-Saclay, Synchrotron Soleil, 91190, Saint-Aubin, France.\\
$^{6}$ Massachusetts Institute of Technology MIT, Department of Chemistry, 77 Massachusetts Ave, Cambridge, MA 02139, USA \\
$^{7}$Laboratoire Structures, Propri\'et\'es et  Mod\'elisation des Solides, CentraleSup\'elec, UMR CNRS 8580, Universit\'e Paris-Saclay, 91190 Gif-sur-Yvette, France}

\maketitle

\section{Supplementary Note 1}

We show in the Supplementary Figure 1, the correspondance between the cubic, the rhombohedral and hexagonal frames. As discussed in the text the plane $(Ox'_2x_3)$, within the hexagonal or trigonal frame, is a symmetry plane for BiFeO$_3$ and corresponds to the $(110)_c$ plane in the cubic approximation. 

\begin{figure}
\centerline{\includegraphics[width=9cm]{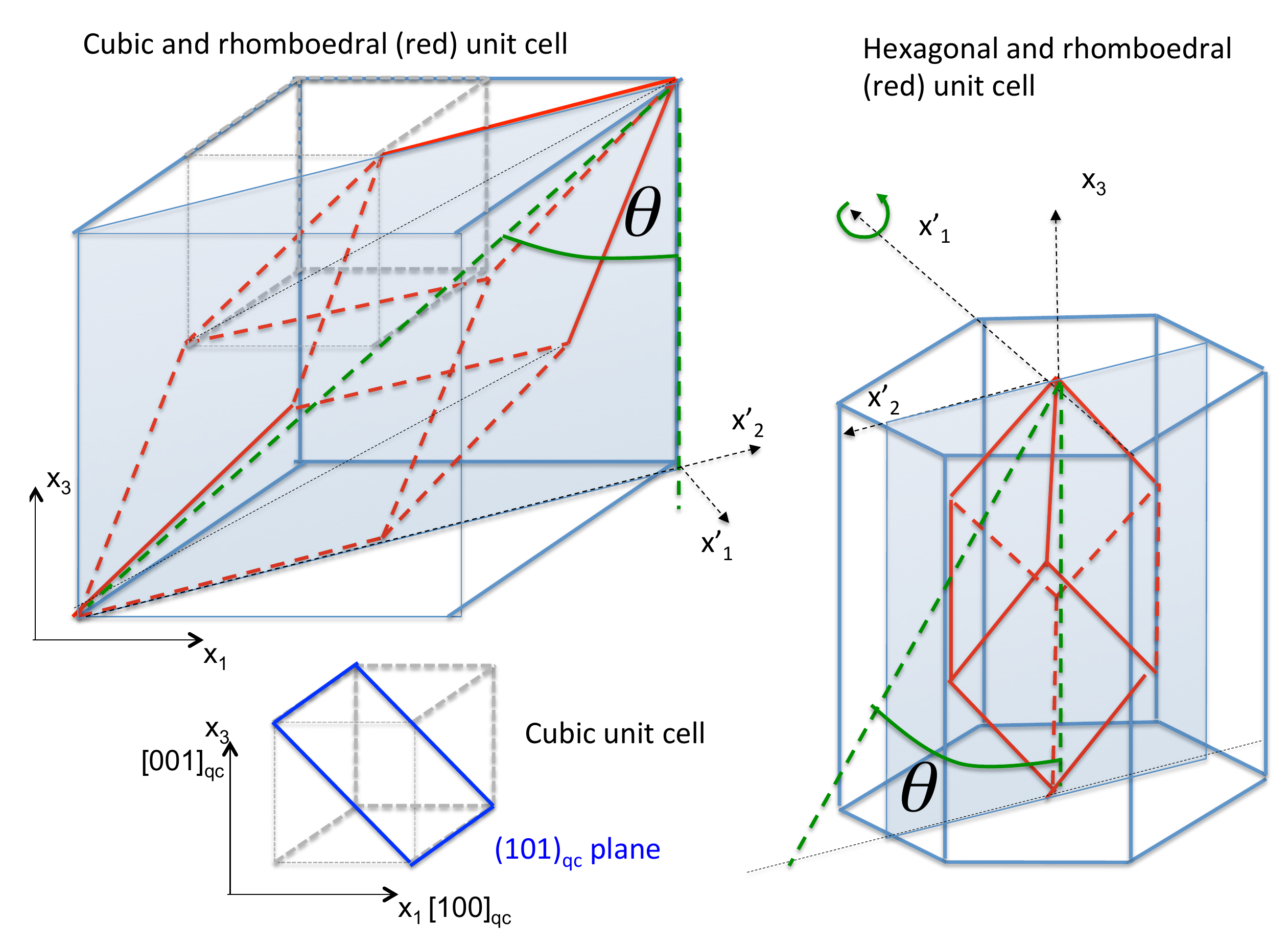}}
\caption{\label{fig1SM} %\textbf{Photonduced longitudinal and shear strain in BiFeO$_3$} 
Cubic, rhomboedral (trigonal) and hexagonal frames. $\theta$ is the angle between the C$_3$ axis of the trigonal BFO structure and the normal of the irradiated surface (free surface) which is along x$_3$. The tensors in the trigonal frame discussed in Supplementary Note 4, are expressed in the coordinates ($Ox'_1x'_2x_3$). }
\end{figure}
\newpage
\section{Supplementary Note 2}

\begin{figure}[t!]
\centerline{\includegraphics[width=7cm]{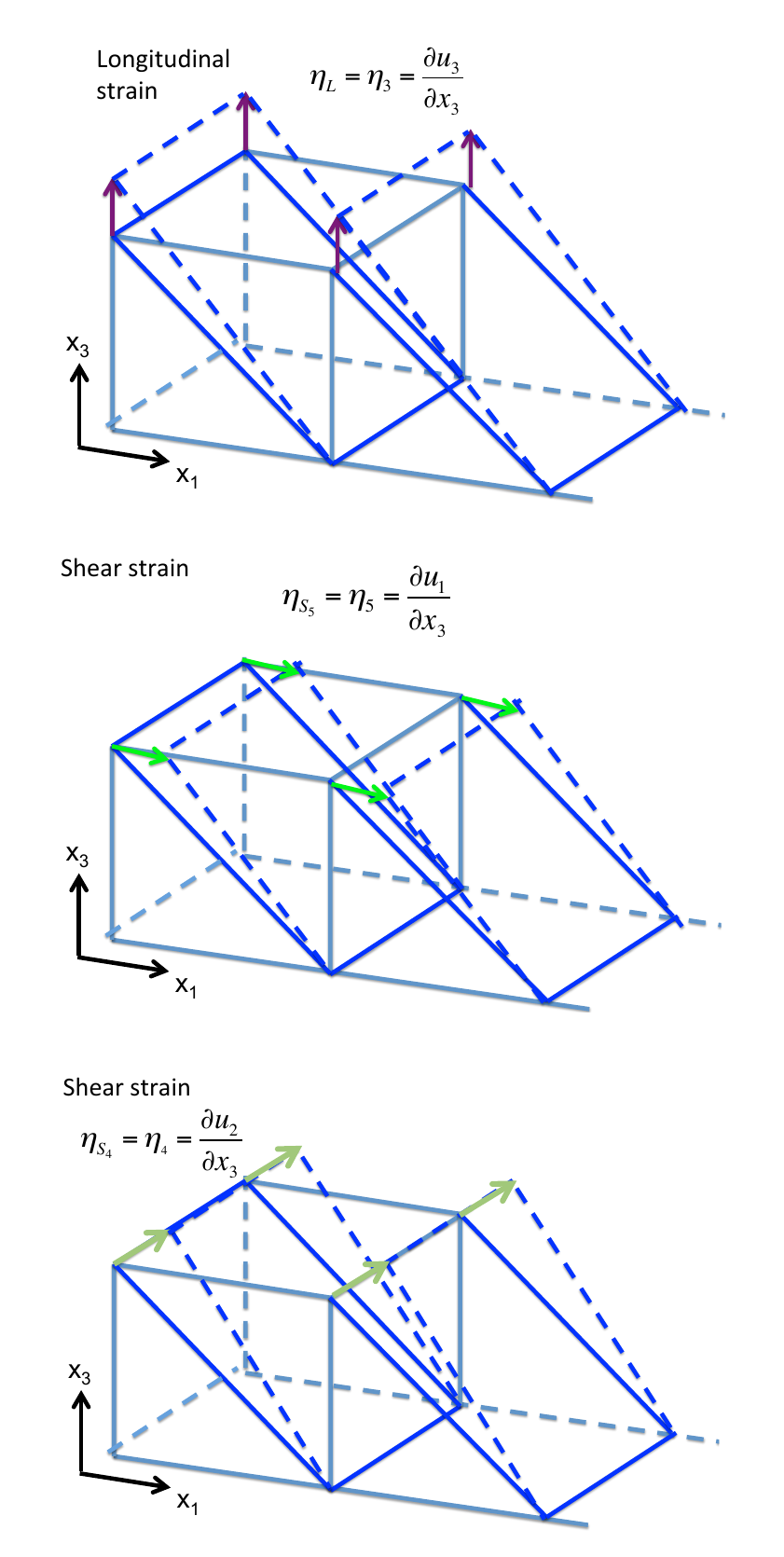}}
\caption{\label{fig2} %\textbf{Photonduced longitudinal and shear strain in BiFeO$_3$} 
Picture of the strained pseudo-cubic unit cell in presence of a longitudinal ($\eta_L=\frac{\partial u_3}{\partial x_3}$) and shear ($\eta_4=\frac{\partial u_2}{\partial x_3}$, $\eta_5=\frac{\partial u_1}{\partial x_3}$) strains. $\theta \sim 54^{\circ}$ is the angle between the C$_3$ axis of the trigonal BFO structure and the normal of the irradiated surface (see Suppl. Fig. 1). }
\end{figure}

In this note we describe all the physical entities in the cubic frame $(Ox_1x_2x_3)$ as shown in Suppl. Fig. \ref{fig1SM}.
The deformation of the cubic unit cell associated to the longitudinal ($\eta_L=\eta_3=\frac{\partial u_3}{\partial x_3}$) and shear  ($\eta_{S_5}=\eta_5=\frac{\partial u_1}{\partial x_3}, \eta_{S_4}=\eta_4=\frac{\partial u_2}{\partial x_3}$) is depicted in the following Supplementary Figure (Fig. \ref{fig2}). We can see that the shear strain $\eta_{S_4}=\eta_4$ does not contribute to the modification of the interplanar distance of the planes (101), since it induces only a sliding motion. 

\begin{figure}[t!]
\centerline{\includegraphics[width=9cm]{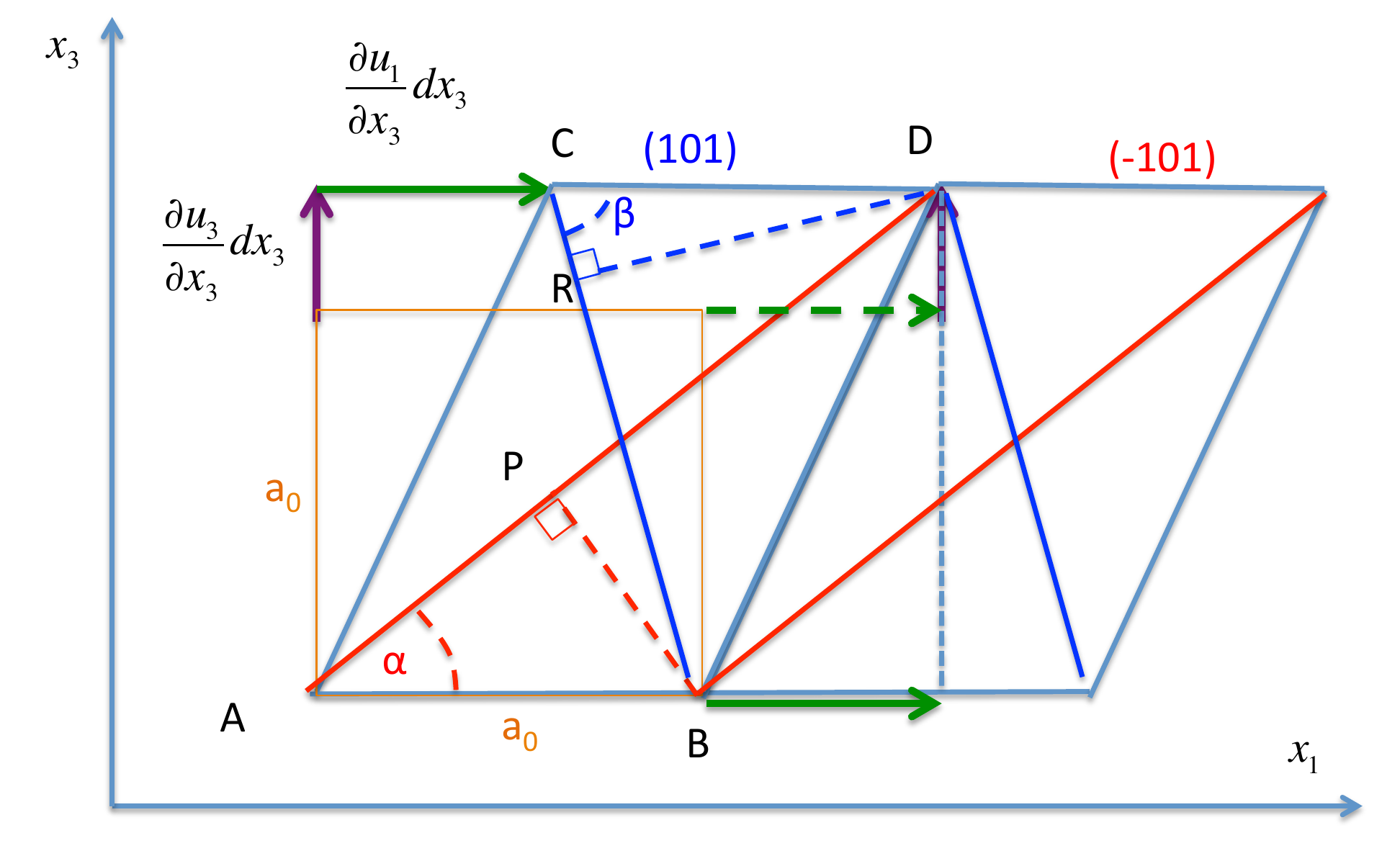}}
\caption{\label{fig3} %\textbf{Photonduced longitudinal and shear strain in BiFeO$_3$} 
2D Picture of the strained cubic unit cell in presence of a longitudinal ($\eta_L=\frac{\partial u_3}{\partial x_3}$) and shear ($\eta_{S_5}=\frac{\partial u_1}{\partial x_3}$) strain.}
\end{figure}

We can then demonstrate how the interplanar distances evolve in presence of strains thanks to the following geometrical calculation (Fig. \ref{fig3}). In this figure, the projection of the unstrained pseudo-cubic unit cell in the $(O,x_1,x_3)$ plane is represented in orange. The pseudo-cubic lattice parameter is denoted $a_0$. The $(101)$ and $(-101)$ interplanar distances in the strained unit cells correspond to the lengths of the $RD$ and $PB$ segments, respectively. We have : 

    \begin{eqnarray}
\label{appen}
PB=\sqrt{{a_0}^2-AP^2}\nonumber\\
AP=a_0 \times cos(\alpha)\nonumber\\
cos(\alpha)=\frac{a_0+\eta_{S_5} a_0}{\sqrt{(a_0+\eta_{S_5}a_0)^2+(a_0+\eta_{L}a_0)^2}}
\end{eqnarray}

with a Taylor development for which $\eta_L, \eta_{S_5} << 1$, we arrive to : 

\begin{eqnarray}
\label{appen}
PB=d^*_{-101} \approx \frac{a_0}{\sqrt2}(1+ \frac{\eta_L-\eta_{S_5}}{2}) 
\end{eqnarray}

Similarly, we can write for $d^*_{101}$=RD:
    \begin{eqnarray}
\label{appen}
RD=\sqrt{{a_0}^2-CR^2}\nonumber\\
CR=a_0 \times cos(\beta)\nonumber\\
cos(\beta)=\frac{a_0-\eta_{S_5} a_0}{\sqrt{(a_0+\eta_{L}a_0)^2+(a_0-\eta_{S_5}a_0)^2}}
\end{eqnarray}

which, with a Taylor development, leads to : 

\begin{eqnarray}
\label{appen}
RD=d^*_{101} \approx \frac{a_0}{\sqrt2}(1+ \frac{\eta_L+\eta_{S_5}}{2}) 
\end{eqnarray}

By combining the above equations and generalizing that situation to the family (h01), we arrive to 

\begin{eqnarray}
\label{appen}
\frac{\Delta d_{\pm h01}}{d_{\pm h01}}=\frac{d^*_{\pm h01}-d_{\pm h01}}{d_{\pm h01}}=\frac{\eta_L\pm \mid h\mid \eta_{S_5}}{1+h^2}
\end{eqnarray}

It has to be noted previously that only $\eta_{S5}$ contributes to the symmetry breaking for the $(\pm h01)$ planes. Considering that the total amplitude of the shear strain is $\eta_S=\sqrt{\eta_{S4}^2+\eta_{S5}^2}=\sqrt{2} \times \eta_{S5}$ as $\eta_{S4}=\eta_{S5}$, then we can write as well : 

\begin{eqnarray}
\label{appen}
\frac{\Delta d_{\pm h01}}{d_{\pm h01}}=\frac{d^*_{\pm h01}-d_{\pm h01}}{d_{\pm h01}}=\frac{\eta_L\pm \mid h\mid \frac{\eta_{S}}{\sqrt{2}}}{1+h^2}
\end{eqnarray}

Consequently we deduce the longitudinal and shear strains as follows:

\begin{eqnarray}
\label{strain}
\eta_L&=&\frac{1+h^2}{2} \times [{\frac{d^*_{h01}-d_{h01}}{d_{h01}}+\frac{d^*_{-h01}-d_{-h01}}{d_{-h01}}}]\nonumber\\
\eta_{S}&=&\frac{1+h^2}{\sqrt{2}\mid h\mid} \times [{\frac{d^*_{h01}-d_{h01}}{d_{h01}}-\frac{d^*_{-h01}-d_{-h01}}{d_{-h01}}}]\nonumber\\
\end{eqnarray}

\section{Supplementary Note 3}

%\begin{figure}[t!]
%\centerline{\includegraphics[width=7cm]{fig4SM.pdf}}
%\caption{\label{fig4} %\textbf{Photonduced longitudinal and shear strain in BiFeO$_3$} 
%Principle of the time-resolved X-ray diffraction with grazing incidence.}
%\end{figure}

The diffraction patterns were recorded with the 2D detector XPAD3.2 whose coordinates ($\gamma, \delta$) are depicted in the scheme given in Figure 2 of the main manuscript. The Bragg angle $\theta$ is deduced from the relation $cos(2\theta)=cos(\delta)cos(\gamma)$. 
%The Bragg peak is recorded in the 2D detector thanks to a kphi scan as shown in Fig. \ref{fig2}(b). 
We have extracted the center of mass (CoM) for each $\phi$ scan to determine $\Delta \gamma(t)$ and $\Delta \delta(t)$ and to finally deduce the relative variation of the interplanar distance variation as :

\begin{eqnarray}
\label{db}
\frac{{\Delta d_{\pm h01}}{(t)}}{d_{\pm h01}}=\frac{1}{2}\frac{cos(\delta)sin(\gamma)\Delta \gamma (t)+ sin(\delta)cos(\gamma)\Delta \delta (t)}{cos(\delta)cos(\gamma)-1}
\end{eqnarray}

where $\frac{\Delta d_{\pm h01}(t)}{d_{\pm h01}}=\frac{d^*_{\pm h01}(t)-d_{\pm h01}}{d_{\pm h01}}$  with $d^*_{\pm h01}(t)$ the interplanar distance after the laser excitation. 
%The time-resolved X-ray diffraction experiments were conducted in a quasi-stationnary regime where the pulsed femtosecond laser continuously impinged on the sample. When the photoexcitation was stopped, we have observed a very slow relaxation process of the lattice with characteristic time of the order of the minute. This very slow dynamics is out of the scope of the paper where we concentrate on the ultrafast response of the lattice after femtosecond light excitation.

\section{Supplementary Note 4}

In this note we describe the properties of the acoustic waves that propagate in the anisotropic BFO crystal after the photoexcitation process. We establish the general expression of the propagating longitudinal and shear strain as a function of out-of and in-plane components of the quasi-longitudinal and quasi-transverses acoustic modes found by solving the Christoffel equation. We describe all the physical entities in the trigonal frame $(Ox'_1x'_2x_3)$ as shown in Suppl. Fig. \ref{fig1SM}. As an example the Voigt component of index $2$ refers to axis $x'_2$. 

%\subsection{Photoelastic tensor}

%The photoelastic tensor of BiFeO$_3$ is a four rank tensor whose elements are not known unfortunately. In our experiment the detection of photoacoustic signals (Fig. 1 in the main manuscript) has been achieved with circular probe polarization, i.e. with probe light having electric field components along the $x'_1$ and $x'_2$ axis. Note that since the trigonal axis is inclined regarding to the normal of the irradiated surface, the internal electric field (within BFO) has an light electric field component along $x_3$). Once the photoelastic tensor is rotated around the $x'_1$ axis, the relevant disturbed dielectric constant tensor ($\Delta [\epsilon]$) in a presence of both longitudinal ($\eta_{33}=\eta_L$) and shear ($\eta_{23}=\eta_S$) strain, becomes then : 

%\begin{eqnarray}
%\label{photoelas}
%\Delta \epsilon_{11}&=&P_{13}\eta_{33}+P_{14}\eta_{23}=P_{13}\eta_{L}+P_{14}\eta_{S} \nonumber\\
%\Delta \epsilon_{22}&=&P_{23}\eta_{33}+P_{24}\eta_{23}=P_{23}\eta_{L}+P_{24}\eta_{S} \nonumber\\
%\Delta \epsilon_{33}&=&P_{33}\eta_{33}+P_{34}\eta_{23}= P_{33}\eta_{L}+P_{34}\eta_{S}\nonumber\\
%\Delta \epsilon_{23}&=&P_{43}\eta_{33}+P_{44}\eta_{23}=P_{43}\eta_{L}+P_{44}\eta_{S} \nonumber\\
%\Delta \epsilon_{13}&=&0 \nonumber\\
%\Delta \epsilon_{12}&=&0 \nonumber\\
%\end{eqnarray}

\subsection{Christoffel equation}

The Christoffel equation permits to describe the displacement vector of the acoustic modes in any direction of a crystal having a given symmetry \cite{royer}. For the particular case of our geometry, for a wave propagating along the $x_3$ direction, the Christoffel tensor becomes : 
\begin{eqnarray}
%\left (
\Gamma_{x_3}=
\begin{pmatrix} C_{55} & 0 & 0\\
0 & C_{44}  & C_{34} \\
0 & C_{34} & C_{33} \\

 \end{pmatrix}
% \right )
\end{eqnarray}

This indicates that a pure shear wave with a particle displacement along $x'_1$ propagating at the sound velocity $V_{TA}=\sqrt{C_{55}/\rho}$ can propagate along $x_3$. But, as mentioned in the main manuscript, this wave is not generated for symmetry reason : the (110)$_{c}$ is a symmetry plane. That plane in the cubic frame corresponds to the so-called ($Ox'_2x_3$) in the trigonal frame (see Supplementary Note 1). There are also two waves (named QLA and QTA) having particle displacement in the plane $Ox'_2x_3$ and by solving the Christoffel determinant  : 

\begin{eqnarray}
%\left (
\Delta=
\begin{vmatrix}  C_{44}-\rho V^2  & C_{34} \\
 C_{34} & C_{33}-\rho V^2 \\
 \end{vmatrix}
 =0
% \right )
\end{eqnarray}
We show that the quasi-longitudinal and a quasi-tranverse modes velocities are:
\begin{eqnarray}
\label{VQLATA}
V_{QLA}=\sqrt{\frac{(C_{44}+C_{33})+\sqrt{(C_{44}-C_{33})^2+ 4C_{34}^2}}{2\rho}}\nonumber\\
V_{QTA}=\sqrt{\frac{(C_{44}+C_{33})-\sqrt{(C_{44}-C_{33})^2+4C_{34}^2}}{2\rho}}
\end{eqnarray}

The values of the elastic stiffness coefficients in the rotated base used to estimate these velocities, are given in the following tensor: 

  \begin{eqnarray}
 C_{ijkl}(\theta=54^{\circ})=
\begin{pmatrix}
207. & 93.291 & 79.709 & -40.5849 & 0. & 0. \\
 93.291 & 130.083 & 77.4306 & 9.93573 & 0. & 0. \\
 79.709 & 77.4306 & 181.056 & -26.8898 & 0. & 0. \\
 -40.5849 & 9.93573 & -26.8898 & 57.4306 & 0. & 0. \\
 0. & 0. & 0. & 0. & 19.784 & -11.5777 \\
 0. & 0. & 0. & 0. & -11.5777 & 52.216 \\
\end{pmatrix}
\end{eqnarray}

The non-rotated tensor values come from the literature. $C_{11}$, $C_{33}$, $C_{44}$ and $C_{66}$ come from experimental values found in Ref. \cite{bori} (note that these authors have used a slightly different density for BFO as ~8.4g.cm$^{-3}$). We remind that $C_{66}=(C_{11}-C_{12})/2$. $C_{14}$ and $C_{13}$ are calculated values (GGA method) found in Ref. \cite{shang}. The non-rotated elastic stiffness tensor is (value in GPa). 
 
 \begin{eqnarray}
 C_{ijkl}(\theta=0)=
\begin{pmatrix}
 207. & 123. & 50. & 19. & 0. & 0. \\
 123. & 207. & 50. & -19. & 0. & 0. \\
 50. & 50. & 159. & 0. & 0. & 0. \\
 19. & -19. & 0. & 30. & 0. & 0. \\
 0. & 0. & 0. & 0. & 30. & 19. \\
 0. & 0. & 0. & 0. & 19. & 42. \\
\end{pmatrix}
\end{eqnarray}
 
Using the elastic stiffness constants in the rotated frame, we arrive to : $V_{QLA} \approx$ 4710 m/s and $V_{QTA} \approx$ 2480 m/s. In our experiments (see Fig. 1 in the main manuscript), we measured $V_{QLA} \approx$ 4970 m/s and $V_{QTA} \approx$ 3020 m/s. The relative difference of around $5-20\%$ is important but is typically the one we find when comparing the different data of the literature \cite{bori,shang}. 

The eigen vectors, i.e. the in-plane and out-of plane components of these QLA and QTA waves are  : $\vec{u}^{QLA} \approx(0,1,4.47)$ and $\vec{u}^{QTA}\approx(0,-4.47,1)$ and are represented in Fig. 3(d) of the main manuscript. The general expressions of the time and space dependence of the particle displacements associated to the QLA and QTA modes become then: 

\begin{eqnarray}
\label{displ}
\vec{u}^{QLA}=A\times f(t-\frac{x_3}{V_{QLA}}) \vec{x}'_2+ 4.47A \times f(t-\frac{x_3}{V_{QLA}})\vec{x}_3 \nonumber\\
\vec{u}^{QTA}=-4.47B \times f(t-\frac{x_3}{V_{QTA}}) \vec{x}'_2+ B \times f(t-\frac{x_3}{V_{QTA}})\vec{x}_3\nonumber\\
\end{eqnarray}

A and B are the amplitudes and $f(t-\frac{x_3}{V_{QLA}})$ and $f(t-\frac{x_3}{V_{QTA}})$ describe the propagative nature of the acoustic waves associated to the QLA and QTA modes respectively.

The associated strains become : 

\begin{eqnarray}
\label{displ}
\vec{\eta}^{QLA}&=&\frac{\partial \vec{u}^{QLA}}{\partial x_3}\nonumber\\
&=& \frac{-A}{V_{QLA}}\times f(t-\frac{x_3}{V_{QLA}}) \vec{x}'_2\nonumber\\
&+& \frac{-4.47A}{V_{QLA}}  \times f(t-\frac{x_3}{V_{QLA}})\vec{x}_3 \nonumber\\
\vec{\eta}^{QTA}&=&\frac{\partial \vec{u}^{QTA}}{\partial x_3}\nonumber\\
&=&\frac{4.47B}{V_{QTA}} \times f(t-\frac{x_3}{V_{QTA}}) \vec{x}'_2\nonumber\\
&+& \frac{-B}{V_{QTA}}  \times f(t-\frac{x_3}{V_{QTA}})\vec{x}_3\nonumber\\
\end{eqnarray}

The total in-plane (shear) and out-of-plane (longitudinal) strains coming from both contributions of the QTA and QLA modes, consist in the addition of components along $x'_2$ and $x_3$ respectively. We have: 

\begin{eqnarray}
\label{strainLS3}
\eta_T(t)&=& \frac{4.47B}{V_{QTA}} f(t-\frac{x_3}{V_{QTA}})+ \frac{-A}{V_{QLA}}f(t-\frac{x_3}{V_{QLA}})  \nonumber\\
\eta_L(t)&=&\frac{-B}{V_{QTA}} f(t-\frac{x_3}{V_{QTA}})+\frac{-4.47A}{V_{QLA}}f(t-\frac{x_3}{V_{QLA}})\nonumber\\
\end{eqnarray}

\subsection{Modeling the photoinduced strain functions}

In the main manuscript we have defined the functions $f(t-x_3/V_{QTA})$ and $f(t-x_3/V_{QLA})$ which describe the time and space dependence of the photoinduced strain  to QLA and QTA modes. As a first very simple approach, we have considered that the emitted coherent acoustic pulse can be derived according to Thomsen et al \cite{tom1} describing the space and time dependence of a bipolar acoustic pulse emitted from the free surface of a material after an ultrafast photoexcitation with the stationary strain near the surface. We have considered this Thomsen's function as representative for $f(t-x_3/V_{QTA})$ and $f(t-x_3/V_{QLA})$ with \cite{tom1}: % but with different amplitude for the QLA and QTA modes. %We have then the associated in-plane ($\eta_S^{QTA}, \eta_S^{QLA}$) and out-of plane ($\eta_L^{QTA}, \eta_L^{QLA}$) components. 
%In this model  \cite{tom1}, the photoinduced stress is considered as instantaneous and takes place within the optical skin depth of pump radiation. We have  \cite{tom1}: 

\begin{eqnarray}
\label{strainLS2}
f(t-x_3/V_{QTA})&=& [e^{-x_3/\xi_{p}}(1-\frac{1}{2}e^{-V_{QTA}t/\xi_{p}})\nonumber\\
&-&\frac{1}{2}e^{-\mid{x_3-V_{QTA}t}\mid/\xi_p}sgn(x_3-V_{QTA}t)] \nonumber\\
f(t-x_3/V_{QLA})&=& [e^{-x_3/\xi_{p}}(1-\frac{1}{2}e^{-V_{QLA}t/\xi_{p}})\nonumber\\
&-&\frac{1}{2}e^{-\mid{x_3-V_{QLA}t}\mid/\xi_p}sgn(x_3-V_{QLA}t)] \nonumber\\
\end{eqnarray}

with $\xi_p \sim 40 nm$ the penetration depth of the pump beam. The general longitudinal and shear strain become:
\begin{equation}
\label{strainLS2}
  \begin{aligned}
\eta_{S}(t,x_3)&=& \eta_S^{QTA} f(t-\frac{x_3}{V_{QTA}}) +\eta_S^{QLA} f(t-\frac{x_3}{V_{QLA}}) \\
\eta_{L}(t,x_3)&=& \eta_L^{QTA} f(t-\frac{x_3}{V_{QTA}}) +\eta_L^{QLA} f(t-\frac{x_3}{V_{QLA}}) ,
  \end{aligned}
\end{equation}
%Since the measured photoinduced strain is a result of an average of the lattice parameter over the X-ray probe penetration ($\xi_X \sim 70 nm$), we have calculated the average strain like, as an example, $ <{f(t-x_3/V_{QTA})}>= \frac{1+F'}{\xi_X}\int_{x_3=0}^{\infty}e^{-x_3/{L_{X}}}f(t-x_3/V_{QTA})dx_3$ ($x_3=0$ corresponds to the free surface of the crystal). $L_X$ is the effective depth over which the average of light-induced strain is realized and $F'$ is a geometric factor taking into account the path of the grazing incidence of the X-ray beam. These parameters are discussed in the next subsection. The calculated shear and longitudinal strains shown with the blue and orange colors in Fig. 3(c) of the main manuscript are described as : 

%\begin{eqnarray}
%\label{strainLS3}
%%\overline{\eta_T(t)}&=&\eta_S^{QTA}\overline{f(t-\frac{x_3}{V_{QTA}})}+\eta_S^{QLA}\overline{f(t-\frac{x_3}{V_{QLA}})}\nonumber\\
%%\overline{\eta_L(t)}&=&\eta_L^{QTA}\overline{f(t-\frac{x_3}{V_{QTA}})}>+\eta_L^{QLA}\overline{f(t-\frac{x_3}{V_{QLA}})}\nonumber\\
%<{\eta_T(t)}>&=&\eta_S^{QTA}<{f(t-\frac{x_3}{V_{QTA}})}>+\eta_S^{QLA}<{f(t-\frac{x_3}{V_{QLA}})}>\nonumber\\
%<{\eta_L(t)}>&=&\eta_L^{QTA}<{f(t-\frac{x_3}{V_{QTA}})}>+\eta_L^{QLA}<{f(t-\frac{x_3}{V_{QLA}})}>\nonumber\\
%\end{eqnarray}
%
%
The parameters $\eta_S^{QTA}, \eta_S^{QLA}, \eta_L^{QTA}$ and $\eta_L^{QLA}$ which depend on the $A$, $B$, $V_{QLA}$ and $V_{QTA}$, are fitted in our Fig. 3d (main manuscript). Note that the parameters $A$ and $B$ depend on the photogeneration processes. %This average strain ( $ <F(t-x_3/V_{QTA})>= \frac{1}{\xi_X}\int_{x_3=0}^{\infty}e^{-x_3/\xi_{X}}f(t-x_3/V_{QTA})dx_3$) is plotted versus time in Fig. 3(c) of the main manuscript (blue and orange solid lines in the  left panel) taking into account the finite probed volume $e^{-x_3/\xi_{X}}$. \\

\section{Supplementary Note 5}

\begin{figure}
\centerline{\includegraphics[width=12cm]{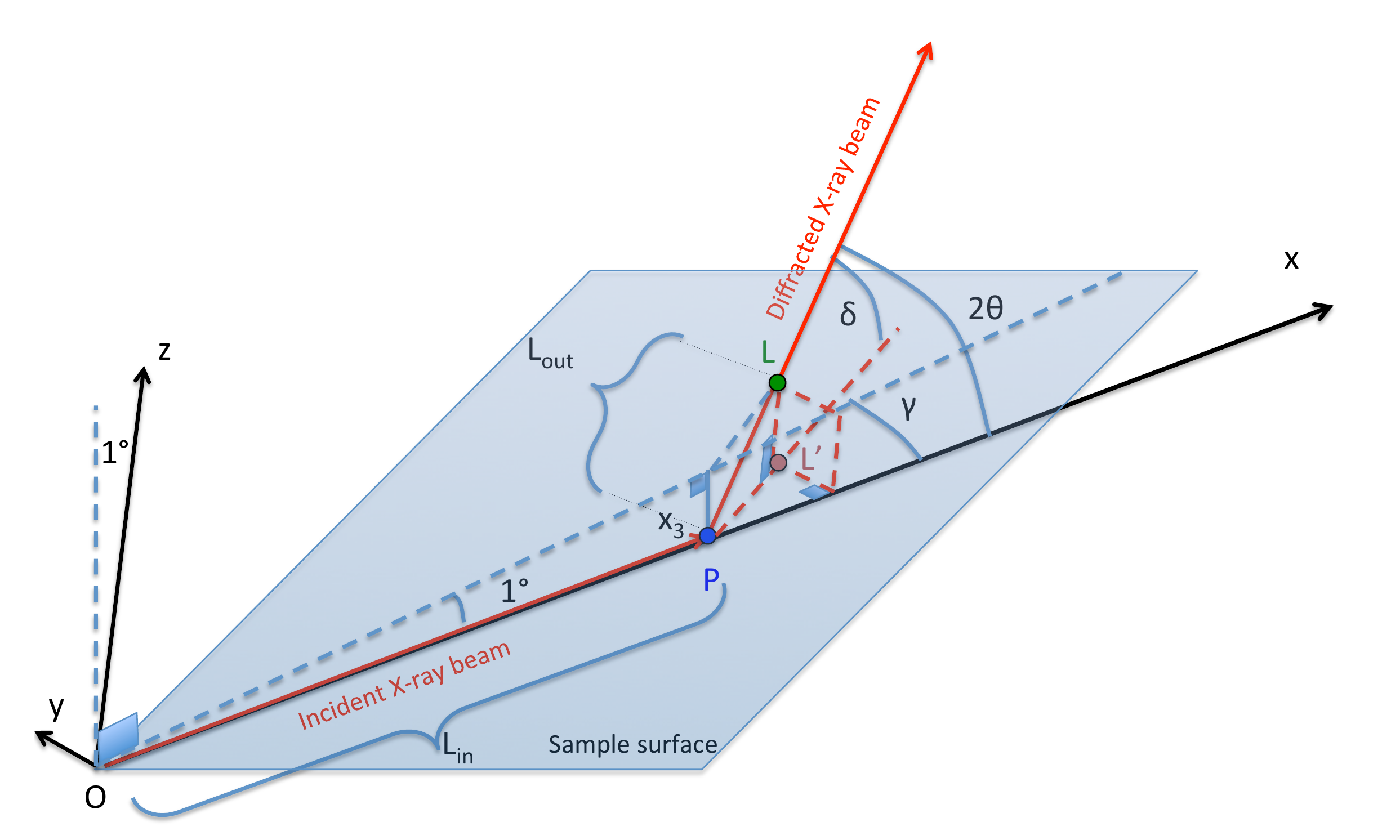}}
\caption{\label{claire} %\textbf{Photonduced longitudinal and shear strain in BiFeO$_3$} 
Calculation of the path of the incident and scattered X-ray beam as a function of the depth $x_3$.}
\end{figure}

In the following, we derive the expression of the length travelled by the X-ray beam within the BFO sample when diffracted at the depth $x_3$. For that we present in the Fig. \ref{claire} the geometry of the diffraction experiment with the incident $L_{in}=[OP]$ and scattered $L_{out}=[PL]$ beam paths. The entrance point of the incident beam is $O$ and the exit point of the scattered beam is $L$. The scattering process takes place at point $P$. The incident and output paths length are:

\begin{eqnarray}
\label{path1}
L_{in}&=&x_3/sin(1^{\circ})\nonumber\\
L_{out}&=&\sqrt{(x_L-L_{in})^2+y_L^2+z_L^2}\nonumber\\
\end{eqnarray}

The coordinates $(x_L,y_L,z_L)$ are defined by the Bragg conditions, i.e. by the angles ($\delta,\gamma$), with: 

\begin{eqnarray}
\label{path2}
y_L&=&(x_L-L_{in})tan(\gamma)\nonumber\\
z_L&=&x_Ltan(1^{\circ})\nonumber\\
\end{eqnarray}

Another expression of $z_L$ can be derived by considering the projection of $[PL]$ in the (O,x,y) plane, denoted $[PL']$:

\begin{eqnarray}
\label{path3}
[PL']^2=[(x_L-L_{in})^2+y_L^2+z_L^2]cos^2(\delta)&=&(x_L-L_{in})^2+y_L^2\nonumber\\
z_L&=&tan(\delta)\sqrt{(x_L-L_{in})^2[1+tan^2(\gamma)]}\nonumber\\
\end{eqnarray}

We need then to find $x_L$ as a function of the experimental parameters. With $z_L=x_Ltan(1^{\circ})$ and considering Eq. \ref{path3}, we arrive to $x_Ltan(1^{\circ})=tan(\delta)\sqrt{(x_L-L_{in})^2[1+tan^2(\gamma)]}$ which, once squared, leads to a second-degree polynomial expression:

\begin{eqnarray}
\label{path4}
x^2_L[tan^2(1^{\circ})-G]+x_L[2L_{in}G]-L_{in}^2G&=&0\nonumber\\
\end{eqnarray}

with $G=tan^2(\delta)[1+tan^2(\gamma)]$. The solution of this equation is:

\begin{eqnarray}
\label{path5}
x_L&=&L_{in}\left({\frac{G+\sqrt{G}tan(1^{\circ})}{G-tan^2(1^{\circ})}}\right)=L_{in}F\nonumber\\
\end{eqnarray}

The total path $L=L_{in}+L_{out}$ becomes:

\begin{eqnarray}
\label{path6}
L&=&L_{in}\left[1+ (F-1)\sqrt{1+tan^2(\gamma)+tan^2(\delta)(1+tan^2(\gamma))}\right]=L_{in}\left[1+ F'\right]=\frac{x_3}{sin(1^{\circ})}\left[1+ F'\right]\nonumber\\
\end{eqnarray}

The $F'$ factor is found equal to 0.0416 for the $(\pm101)$ Bragg peaks and equal to 0.0407 for the $(\pm201)$ Bragg peaks.

\section{Supplementary Note 6}

In this section, we describe the theoretical calculation of the photoinduced stresses used in Eq. 5 and 6 of the main manuscript. In the photoexcitation process, the equation of motion must include the source term, i.e. the photoinduced shear $\sigma_2$ (source term for the in-plane displacement $u_2$ along $x'_2$) and longitudinal $\sigma_3$ (source term for the out-of-plane displacement $u_3$ along $x_3$) according to: 
\begin{eqnarray}
\label{strainLS2}
\rho \frac{\partial^2u_2(x_3,t)}{\partial t^2}-C_{44}\frac{\partial^2u_2(x_3,t)}{\partial x_{3}^2}-C_{43}\frac{\partial^2u_3(x_3,t)}{\partial x_{3}^2}=\frac{\partial \sigma_2(x_3,t)}{\partial x_{3}} \nonumber\\
\rho \frac{\partial^2u_3(x_3,t)}{\partial t^2}-C_{43}\frac{\partial^2u_2(x_3,t)}{\partial x_{3}^2}-C_{33}\frac{\partial^2u_3(x_3,t)}{\partial x_{3}^2}=\frac{\partial \sigma_3(x_3,t)}{\partial x_{3}} \nonumber\\
\end{eqnarray}

These photoinduced stresses can then be described by the inverse piezoelectric or the thermoelastic effects, or by a combination of both. (Remark : without the source term we come back to the Christoffel equation, see Supplementary Note 4)

We have established in a previous paper the general expression of the photoinduced stress driven by the piezoelectric effect  \cite{lejman2014} with : 

\begin{eqnarray}
\label{piezo}
\sigma_T^{PE}=\sigma_{23}^{PE}=(d_{31}-d_{33})cos(\theta)sin(\theta) E  \nonumber\\
\sigma_L^{PE} =\sigma_{33}^{PE}=(d_{31} cos^2(\theta)+d_{33}sin^2(\theta)) E\nonumber\\
\end{eqnarray}

$d_{31} \sim$ -30/-20  pm.V$^{-1}$ \cite{muri,sichel} and $d_{33} \sim$ 50/70 pm.V$^{-1}$ \cite{chen} are the piezoelectric coefficients. $\theta \sim$ 54$^{\circ}$ and $E$ is the internal depolarizing field.  

The theoretical contribution of the photoinduced thermoelastic stress (TE) $\sigma_{ij}^{TE}$ is assessed by calculating the tensorial expression $\sigma_{ij}^{TE}=C_{ijkl}\beta_{kl}\Delta T$. For that we have to take into account the anisotropic thermal expansion of BiFeO$_3$. The (001)$_{qc}$ cristal  corresponds to a trigonal crystal for which the [001]$_{R}$ direction is rotated by an angle of $\theta$=54$^\circ$ around the $(Ox'_1)$ axis (See Supplementary Figure 1). We remind that $R$ means the trigonal coordinates. Doing such a rotation, the relevant thermal expansion coefficients become  $\beta_4=\beta_{23}=(\beta_a-\beta_c)cos(\theta)sin(\theta)$, $\beta_2=\beta_{22}=\beta_a cos^2(\theta)+\beta_c sin^2(\theta)$ and $\beta_3=\beta_{33}=\beta_a sin^2(\theta)+\beta_c cos^2(\theta)$, with $\beta_a$=0.7.10$^{-5}$K$^{-1}$ and $\beta_c$=1.2.10$^{-5}$K$^{-1}$ the thermal expansion coefficients in the [100]$_{R}$ and [001]$_{R}$ directions at 300K \cite{polonica}. We have also rotated the elastic stiffness tensor \cite{lejman2019} to obtain the final longitudinal ($\sigma_{33}^{TE}$) and shear ($\sigma_{23}^{TE}$) thermoelastic stress components. The rotated tensor has already been given in the Supplementary Note 4. The light-induced thermoelastic shear and longitudinal stresses are then: 

\begin{eqnarray}
\label{thermoelas}
{\sigma_{T}}^{TE}={\sigma_{23}}^{TE}=(C_{41}\beta_{1}+C_{42}\beta_{2}+C_{43}\beta_{3}+2C_{44}\beta_{4})\Delta T\nonumber\\
{\sigma_{L}}^{TE}={\sigma_{33}}^{TE}=(C_{31}\beta_{1}+C_{32}\beta_{2}+ C_{33}\beta_{3}+2C_{34}\beta_{4})\Delta T\nonumber\\
\end{eqnarray}

As mentioned in the main manuscript, looking at the ratio $\sigma_T/\sigma_L$ permits to get rid of the internal depolarizing field $E$ and of the lattice temperature increases $\Delta T$, which both are not known precisely.

\newpage